\documentstyle[12pt,epsf]{article}
\begin{document}
%
%
\newcommand{\nc}{\newcommand}
\nc{\beq}{\begin{equation}}
\nc{\eeq}{\end{equation}}
\nc{\beqa}{\begin{eqnarray}}
\nc{\eeqa}{\end{eqnarray}}
\nc{\lra}{\leftrightarrow}
\nc{\lsim}{\mbox{\raisebox{-.6ex}{~$\stackrel{<}{\sim}$~}}}
\nc{\gsim}{\mbox{\raisebox{-.6ex}{~$\stackrel{>}{\sim}$~}}}
\def\dr{{\rm d}}
\def\ir{{\rm i}}
\def\nr{{\rm n}}
\def\nue{{\nu_e}}
\def\num{{\nu_\mu}}
\def\nut{{\nu_\tau}}
\def\bnue{{{\bar \nu}_e}}
\def\bnum{{{\bar \nu}_\mu}}
\def\bnut{{{\bar \nu}_\tau}}
\def\nutm{{\nu_{\tau -}}}
\def\nutp{{\nu_{\tau +}}}
\def\nutl{{\nu_{\tau \lambda}}}
\def\nul#1{{\nu_{\lambda #1}}}
\def\bnul#1{{{\bar \nu}_{\lambda #1}}}
\def\nutlp{{\nu_{\tau \lambda '}}}
\def\bnutm{{\bar{\nu }_{\tau -}}}
\def\bnutp{{\bar{\nu }_{\tau -}}}
\def\bnutl{{\bar{\nu }_{\tau \lambda}}}
\def\bnutlp{{\bar{\nu }_{\tau \lambda '}}}
\def\numm{{\nu_{\mu -}}}
\def\nump{{\nu_{\mu +}}}
\def\bnumm{{\bar{\nu }_{\mu -}}}
\def\bnump{{\bar{\nu }_{\mu -}}}
\def\mnm{{m_{\nu_\mu}}}
\def\mnt{{m_{\nu_\tau}}}
\def\tg{{T_\gamma}}
\def\Tqcd{{T_{\rm QCD}}}
\def\T100{{T^{100}_{\rm QCD}}}
\def\rqcd{{r_{\rm QCD}}}
\def\nn{{\nonumber}}
\def\dN{{\Delta N}}
\def\dNu{{\dN_\nu}}
\def\dNum{{\dN_\nu^{\rm m}}}
\def\4He{{^4He}}
\def\vs#1{\langle v_{\rm M\o l}\sigma #1 \rangle}
\def\fnstyle{\baselineskip8pt \font \bx cmr8 \bx}
\def\he#1{\hbox{$^{#1}{\rm He}$}}
\def\li#1{\hbox{$^{#1}{\rm Li}$}}
%
%
\begin{titlepage}
\pagestyle{empty}
\baselineskip=18pt
\rightline{CERN-TH/95-335}
\rightline{UMN-TH-1417/95}
\rightline{hep-ph/9512321}
\rightline{December, 1995}
\baselineskip=21pt
\vskip .1in
\begin{center}
{\large{\bf  Nucleosynthesis Limits on the Mass of \\
             Long Lived Tau and Muon Neutrinos}}
\end{center}
\vskip .15truecm
\begin{center}
Brian D.\ Fields,

{\it Department of Physics, University of Notre Dame}

{\it Notre Dame, IN 46556, USA,}

Kimmo Kainulainen{\footnote{On leave of absence from the Department
of High Energy Physics (SEFL), University of Helsinki, Finland.}}

{\it CERN, CH-1211, Geneve 23, Switzerland,}

and

Keith A.\ Olive

{\it School of Physics and Astronomy, University of Minnesota}

{\it Minneapolis, MN 55455, USA.}

\end{center}

\vskip 0.05in
\centerline{ {\bf Abstract} }
\baselineskip=18pt
\vskip 0.5truecm
\noindent
We compute the nucleosynthesis bounds on the masses of stable
Dirac and Majorana neutrinos by solving an evolution equation network
comprising of all neutrino species which in the Dirac case includes
different helicity states as separate species.
We will not commit ourselves to any particular
value of the nucleosynthesis bound
on effective number of light neutrino degrees of freedom $N_\nu$,
but present all our mass bounds as {\em functions} of
$\dNu$. For example, we find that the excluded region in the
mass of a Majorana $\mu$- or $\tau$- neutrino,
$0.31$ MeV $< m_{\nu}^M < 52$ MeV
corresponding to a bound $\dNu < 0.3$ gets relaxed to $0.93$ MeV
$< m_{\nu}^M < 31$ MeV if $\dNu < 1.0$ is used instead.
For the Dirac neutrinos this latter constraint gives the
upper limits (for $T_{\rm QCD} = 100$ MeV): $\mnm < 0.31$ MeV and
$\mnt < 0.37$ MeV.  Also, the constraint $\dNu <1$ {\it allows} a stable
Dirac neutrino with $m^D_{\nu_\tau} > 22$ MeV.

\end{titlepage}
%
%
\newpage
\baselineskip=20pt
\textheight8.3in\topmargin-0.0in\oddsidemargin-.0in

\section{Introduction}

Primordial nucleosynthesis considerations have become a widely used tool
to obtain limits on particle properties such as masses, couplings and
lifetimes.  Nucleosynthesis bounds arise from the tight agreement
between primordial abundances of the light elements deduced from the
observations and the theoretical predictions based on the standard
big bang nucleosynthesis model (SBBN) \cite{wssok}.
Typically, any extension of the
standard model, such as admitting large neutrino masses, could destroy
the agreement between the theoretical prediction and the observational
evidence.

One of the quantities already studied
in the context of nucleosynthesis is the mass of a stable neutrino
\cite{kolb}-\cite{DoTu}. Nucleosynthesis is sensitive
to neutrino masses in the interval $m_\nu \simeq 0.1 - 50$ MeV, and
the actual values of the bounds depend on the particular value
adopted for the nucleosynthesis bound on the effective number
of neutrino degrees of freedom $\dNu$. The nucleosynthesis
bound on $\dNu$ has proven to be difficult to pin down accurately
and has been under constant revision over the past years \cite{wssok},
\cite{OldB}-\cite{FO}.
Most recently, doubts have been raised regarding the consistency of the
standard big bang nucleosynthesis model with 3 massless neutrinos
\cite{Hata}, inducing a closer look into the issue
of possible systematic errors \cite{Copi} in the determination
of element abundances from the observations as well as in the
assumed models of chemical evolution of the light element abundances,
most importantly those of D and \he3 \cite{FO}.
The actual value of the bound, expressed in terms
of $\dNu$, has therefore become harder to evaluate.  Moreover, because the
present experimental upper bound on the tau-neutrino
mass, $m_{\nu_\tau} < 24$ MeV
\cite{Lab0}, is relatively close to the upper end of the hitherto
quoted nucleosynthesis bounds, it would be useful to see how
the nucleosynthesis can compete with the laboratory when
the limit on $\dNu$ is considerably weakened.

The main purposes of this paper are (1) to present a treatment
that is accurate enough in all the subtleties of computation
so that essentially
the only uncertainty in the mass bounds arises from the
abovementioned inherent uncertainty in obtaining
neutrino flavor limits from matching the SBBN predictions
to the observations, and (2) to be general, which is why we will
present our bounds as {\it functions} of the actual nucleosynthesis
constraint.  We will thus always explicitly write down our cross
sections and carefully show how we perform the thermal averages.
Our evolution equations are written in terms of so
called pseudo-chemical potentials $z_i(t)$ \cite{BBF,DK},
and assume only kinetic equilibrium.  This formalism allow us to
follow the evolution of the phase space distribution functions instead
of the integrated number densities and therefore accurately compute the
relevant thermodynamic quantities, such as the energy - or entropy
densities of $\nu$'s. An exception to this is the case of light
Dirac neutrinos, for which the assumption of kinetic equilibrium
does not hold.

We will include all neutrinos in our
equation network. Tracking $\nu_e$ is particularly important, because
$\nu_\tau{\bar \nu_\tau}$  annihilations below the $\nu_e$ decoupling
temperature  $T_{\nu_e} \simeq 2.3$ MeV \cite{EKS},
would produce an excess of $\nu_e$'s around
the $n/p$-freeze out temperature, biasing the $n\leftrightarrow p$
-equilibrium, hence leading to more neutrons being destroyed and
therefore to less helium being produced.
This effect is very large for $m_\nu \sim few$ MeV and
affects our bounds significantly if $\dNu \gsim 1$.

In the Dirac case we treat the different helicity populations  of the
tau neutrino as separate species. Again, the physical reason for
this is simple: for moderately light tau neutrinos the freeze-out
temperature is close to the mass scale, so that $\nut$'s annihilate while
semi-relativistic. Due to the chirality of the interaction, positive and
negative
helicity states interact with different strengths at freeze out, and
therefore can have  different freeze-out number densities compared to what
one finds when assuming averaged interaction strengths and a total equilibrium
between helicity populations \cite{kolb,DR,DoTu}. Somewhat surprisingly,
while each has a large effect on $N_\nu$,
they compensate each other quite
accurately,
so as to give final total abundance in good accordance with the
helicity averaged approach.

Our final results in the Majorana case agree with
some of the earlier results \cite{Kawa,DoTu}, when
restricted to the specific values of the bound on $\dNu$.
In the Dirac case we find stronger upper limit on the disallowed
mass region than ref.\ \cite{kolb} but weaker than that of
ref.\ \cite{DR}.  The upper limits of the excluded regions are
particularly sensitive to the changes in $\dNu$, opening up
a window for a stable tau neutrino below the experimental
bound of 24 MeV if the nucleosynthesis bound is relaxed
to $\dNu > 1.3$ in the Majorana and $\dNu > 0.8$ in the Dirac
cases respectively. The latter possibility is quite plausible.
For the muon neutrino, on the other hand, our upper
limits can be competitive with the
laboratory bound on mass of $\mnm <160$ KeV
\cite{Lab1} only for rather restrictive values of
$\dNu < 0.13$ in the Majorana, and $\dNu \lsim 0.39-0.44$
in the Dirac case.

In section 2 we will derive generic evolution equations for the particle
distribution functions expressed in terms of the pseudo-chemical potential.
We will also discuss some subtleties of incorporating the time-temperature
relationship into the evolution equations.
In section 3 we discuss how the observational bounds on the helium
abundance should be converted into a bound on $N_\nu$. In section 4
we consider the Majorana case and in the section 5
we will derive and solve the equation network with separate equations for
the $\nu_\tau$ helicity components in the Dirac case.
We will pay special attention to the thermal averaging of the helicity
amplitudes, which is a nontrivial task because of the lack of the Lorentz
invariance of the spin dependent matrix elements \cite{DKR}.
Finally, section 6 contains our conclusions. Some calculational
details are presented in the appendix.
%

\section {Generic evolution equations}

In this section we will derive the evolution equations for the particle
distribution functions. We will also discuss how the time-temperature
relation should be consistently incorporated to the equation network.
Our derivation here relies heavily on that of ref.\ \cite{DK}.
We begin by writing down a set of Boltzmann equations for the
scalar phase space distribution functions:
\beq
E_i(\partial_t + pH\partial_p)f_i(p,t) = C_{{\rm E},i}(p,t)
+ C_{{\rm I},i}(p,t),
\label{bolzmann1}
\eeq
where $E_i = (p^2 + m_i^2)^{1/2}$ and
$H = (8\pi\rho/3 M_{\rm Pl}^2)^{1/2}$ is
the Hubble expansion rate, where $M_{\rm Pl}$ is the Planck mass and
$\rho$ is the total energy density. The index $i$ runs over all particle
species in the plasma; each momentum state in each species
has its own equation like (\ref{bolzmann1}), all of which are coupled
together through the elastic and inelastic collision terms
$C_{\rm E}(p,t)$ and $C_{\rm I}(p,t)$.

A tremendous simplification results if the system is in thermal
equilibrium; then each distribution function can be described by
two parameters, the temperature, and possibly, a chemical potential.
Decoupling particle species however, are by definition {\it not} in
thermal equilibrium.  Fortunately though, they often are
in close kinetic equilibrium, because the kinetic equilibrium is
held by elastic scattering processes whose rate typically
greatly surpasses that of annihilations, particularly for large $m/T$.
Therefore, at each instant of time, the momentum distribution of particles
should be closely approximated by a function
\beq
f(p,z_i) \equiv (e^{\beta E_i + z_i}+1)^{-1},
\label{fansaz}
\eeq
where the time dependent function $z(t)$ acts as an effective chemical
potential driving the system out of the chemical equilibrium.
The function $z(t)$ is called pseudo-chemical potential, because,
unlike the ordinary chemical potential, it appears
with the same sign in both the distribution function for particles and
antiparticles \cite{BBF}.

We will assume throughout that photons and electrons, because of their
extremely fast electromagnetic interactions, are in complete thermal
equilibrium and therefore we need not write down evolution equations for
them. For all neutrino species on the other hand it is necessary to follow
the chemical evolution accurately. In the case of muon and tau neutrinos
this is obvious, because it is exactly the effect of their energy density
on the expansion rate, and thereby on the final helium abundance that
we wish to study.
The electron neutrino is known to be nearly massless, so that small
changes in the $\nue$ number would be of no likely importance for the
expansion rate. However, even very small variations in $n_\nue$ are
important because of their direct effect weak reaction rates,
such as $\nue + n \leftrightarrow e^- + p$ that govern the
freezeout of these weak interaction rates and the n/p ratio.
Using the ansatz (\ref{fansaz}) we therefore end up with the
following equations for the various neutrino pseudo-chemical potentials
\beqa
\dot n_i+3Hn_i & = & \sum_{ \{ \alpha \}_i}
               C_\alpha (ij \leftrightarrow kl) \nn \\
               &\vdots & ,
\label{geneq}
\eeqa
where $i$ runs through all neutrino species and the sum
$\{ \alpha \}_i$ is over all the relevant collision channels.
The compactly written left hand sides of the equations are in fact
functions of $z_i$:
\beqa
\dot n_i+3Hn_i &=&  \frac{T_i^3}{2\pi^2}\Biggl\{
                   H(J_1(x_i,z_i) - x_i^2J_{-1}(x_i,z_i) )  \nn \\
               & & \phantom{Ha}
                 + \frac{\dot T_i}{T_i} J_1(x_i,z_i)
                 - \dot \tg J_0(x_i,z_i) \frac{\dr z_i}{\dr \tg }
                                        \Biggr\},
\label{lhs}
\eeqa
where the dot refers to time derivative,
$\tg$ is the photon temperature, $T_i$ is the temperature of
the particle species $i$, $x_i \equiv m_i/T_i$, and the functions
$J_n(x,z)$ are defined as
\beq
J_n(x,z) \equiv \int_0^\infty \dr y \; y^2
(x^2+y^2)^{n/2}\frac{e^{\sqrt{ x^2+y^2}+z}}
{(1+e^{\sqrt{ x^2+y^2}+z})^2}.
\label{Js}
\eeq
We cannot write the right hand sides of (\ref{geneq}) in terms of the
number densities $n(z_i)$ unless we further approximate the phase
space Fermi-Dirac distributions (\ref{fansaz}) with Maxwell-Boltzmann
distributions. This additional approximation has often been made in
deriving relic abundances of decoupling species \cite{kolb,DR,Many,GG}.
We will here keep the more correct FD-statistics and postpone
writing explicit expressions to the collision integrals to the following
sections where particular cases are considered. Note however that we dropped
the contribution from the elastic collision integral, which vanishes under
the assumption of kinetic equilibrium.

The evolution equations (\ref{geneq}) are strongly coupled not only through
the collision terms, but also because of the time-temperature relation; this
is particularly explicit in the form (\ref{lhs}) for the collision part
of the equations (\ref{geneq}). This complication is a general
consequence of the assumption of kinetic equilibrium.
The usual approach to define the time-temperature relationship
(which we will find inadequate) is to assume
that the energy momentum tensor has the particularly simple form
$T^{\mu \nu}= diag(\rho,-p-p-p)$, corresponding to the ideal fluid
approximation, after which the Einstein equations directly lead into
the ``energy conservation'' equation
\beq
\dot \rho = -3H(\rho +p),
\label{econs}
\eeq
where $\rho$ is the total energy density and $p$ is the total pressure.
When energy density and pressure are expressed in terms of integrals
over particle distributions, equation (\ref{econs}) turns into an
additional equation relating time and the photon temperature.

The appearance of other time derivatives, like $\dot{T}_i$ in (\ref{lhs})
arises from our choice to parameterize each distribution function by two
variables: $z_i$ and $T_i$. Complete determination of the evolution
of a system consisting of $N$ separate species would therefore require
$2N+1$ independent equations. It would be possible to obtain additional
collision equations to augment (\ref{econs}), for example by probing
higher moments of the original equation. We will instead find it sufficient
to follow the simplest physical intuition and assume that the neutrino
temperatures are given by the photon temperature down to the scale where
the electrons begin to annihilate, and later follow the reference
temperature of a completely decoupled massless species. That is:
\beq
T_{\nu_i} \equiv \left( \frac{4+2h_e(\tg )}{11}\right)^{1/3}\:\tg
\label{tneu}
\eeq
where the function $h_e$ is related to the electron entropy density
by $s_e = (2\pi^2/45)h_e \tg^3$. This approach becomes better
warranted {\em a posteriori} when we find out that the annihilations
are always practically complete at temperatures well
above the electron annihilation temperature $T_{\rm ann} \simeq m_e/3$.

However, even with a well defined closed set of equations, there
is a problem with the direct use of the na\"{\i}ve energy conservation law
(\ref{econs}).  This has to do with the breakdown of the fundamental ideal
fluid assumption when dealing with an expanding fluid of a nonrelativistic
decoupling species. Indeed, it is known \cite{Weinberg,Bernstein}
that in such systems the energy momentum tensor acquires new terms such
as bulk viscosity.  Neglecting these contributions, by sticking to
the expression (\ref{econs}), eventually leads to a blowup of the time
temperature relation when the energy density in the decoupling species
starts to dominate over the rest of the matter/radiation in the universe.
This only happens at very small temperatures, of course, and including
bulk viscosity terms would exactly cancel the problematic terms
($\dot z_i$s) in (\ref{econs}). The final result of this analysis is
that the intuitive approach works well: namely, the
 photon temperature, to a very good approximation
evolves as a function of time such that the effect of the decoupled
species is only felt through their contribution to the total energy
density (in the Hubble expansion rate). Some
 straightforward algebra based
on this assumption then immediately gives the standard formula
\beq
\frac{\dot \tg}{\tg} =
  -H/\left( 1+\frac{\tg}{3h_I(\tg )}\frac{\dr h_I(\tg )}{\dr \tg}\right),
\label{ttemp}
\eeq
where the function $h_I$ is related to the entropy of the interacting
species, $s_I \equiv (2\pi^2/45)h_I\tg^3$. Combined with equations
(\ref{tneu}) and (\ref{ttemp}) the equations (\ref{geneq}) provides a
consistent set of equations as the starting point of our analysis.
%

\section {Nucleosynthesis constraints in terms of $\dNu$}

Let us now outline the procedure that leads to the nucleosynthesis
constraints on new particle physics models, spelled out in terms of
a bound for the effective number of neutrino species $\dNu = N_\nu - 3$.
The argument goes roughly as follows: whatever the nature of the
new physical phenomenon, its effect on nucleosynthesis eventually
boils down to some calculable change in the primordial helium abundance
$Y_{^4He}$. Since the helium abundance on the other hand is known to be
a monotonic function of energy density of the universe, this change
in $Y_{^4He}$ can be mapped to an {\it effective} change in the energy
density, which customarily is measured in units of energy density
corresponding to one massless neutrino species.

The most stringent nucleosynthesis bounds on arbitrary model parameters are
obtained if one assumes nothing of the likelihood of the underlying
microscopic theory.  Consider the standard model with $N_\nu$
massless neutrinos as a `reference
theory' which will correspond to some unknown extension of standard model.
The connection is made at each value of
the baryon to photon ratio $\eta = n_B/n_\gamma$, in such a way that
in the extended model, the value of the $^4$He abundance,  $Y(\eta)$
is matched at the same value of $\eta$ to a value of $N_\nu$ in the standard
model with the same value of $Y$. This mapping thus has a slight, but
eventually negligible dependence on $\eta$, for a restricted
but relevant range in $\eta$.
Next one computes the likelihood function for the distribution for
the variable $N_\nu$ by comparing BBN predictions with varying $N_\nu$
to the data \cite{OldB,medo,ost}. For example in \cite{ost} this was
found to lead to the best fit
\beq
N_\nu = 2.2 \pm 0.3 \pm 0.4,
\label{Nlim1}
\eeq
which shows the statistical (from the observational determination of
$Y$ and the neutron mean life) and systematic uncertainties (from
$^4$He and to a smaller extent from $\eta$ - in (\ref{Nlim1}), it was
assumed that $\eta_{10} = 3.0 \pm 0.3$).
Since one could well imagine theories that would effectively {\em
lower} the value of $N_\nu$ as well as increase it, one has to, in
the broadest sense we are discussing now, take the bounds (\ref{Nlim1})
seriously, and accept that they might show {\em preference} for some
extension of the standard model predicting less helium.
Based on this
information, the 95\% CL limit was found to be $N_\nu < 3.1$ \cite{ost}.

Systematic errors in the process of inferring the primordial
abundances from the observations however, are not negligible.
The tightest constraints on SBBN for a long time made essential use of the
inferred upper bound on primordial D+\he3-abundance (giving a tight
lower bound on $\eta $); this constraint was utilized also in arriving
(\ref{Nlim1}) \cite{ost}. It has recently been question as to
whether or not these abundances
are subject to particularly large systematic uncertainties due to
their poorly known chemical evolution \cite{FO}. Indeed, because both
chemical and stellar evolution affect the abundances of $^3$He, the uncertainty
is compounded.  Standard stellar models predict that low mass stars will
be efficient producers of $^3$He \cite{it}, a claim which is seemingly backed
up by observations of $^3$He in planetary nebulae \cite{rood}.
However, it appears that when the $^3$He yields are included in
simple models of galactic chemical evolution, no value of $\eta$ leads
to concordance with the observed solar and present abundances of D and
$^3$He \cite{orstv}. The likely preliminary conclusion is that something is
wrong
with the ``standard" models of either chemical and/or stellar evolution as
they pertain to $^3$He.

 Relying only on the much
more robust \he4 and \li7 abundances leads to a shift downwards in the
concordance region for $\eta$, and hence to a distribution that peaks much
closer to $N_\nu = 3$.
Simply taking the observations of \he4 and \li7 at face value,
i.e. without assuming that the systematic errors are
particularly large to artificially produce concordant values of $\eta$, the
combined likelihood functions for \he4 and \li7 show a peak at
$\eta_{10} \equiv 10^{10}\eta = 1.8$ with a 68\% CL range of 1.6 -- 2.8
and a 95\% CL range of 1.4 -- 3.8. This range for $\eta$ can be translated
into a most likely value for $N_\nu = 2.9$.  In fact the analog of eq.
(\ref{Nlim1}) becomes \cite{FO2}
\beq
N_\nu = 3.0 \pm 0.3 \pm 0.4 {~}^{+ 0.1}_{- 0.6},
\label{Nlim2}
\eeq
showing no particular preference to $N_\nu < 3$ (in fact preferring the
standard model result of $N_\nu = 3$) and leading to $N_\nu < 4.0$
at the 95 \% CL level (when adding the errors in quadrature).
 In (\ref{Nlim2}), the first set of errors are the
statistical uncertainties primarily from the observational determination
of $Y$ and is identical to the one in (\ref{Nlim1}). The second set of errors
is the systematic uncertainty arising solely from $^4$He, and the last
set of errors from the uncertainty in the value of $\eta$ and is determined
by the combined likelihood functions of  \he4 and \li7.

However, in light of the problems in treating the systematic
errors, one might rather take a different approach \cite{OS}. Here one
{\em assumes} the correctness of the standard
BBN theory and {\em restricts} ones
scope of extended or modified
theories to only those one is deriving bounds upon.
These new theories have their own prediction of the \he4 abundance,
or possibly a range of predictions corresponding to the
range of acceptable values in their free parameters.
These new \he4 predictions always correspond to effectively having, say,
$N_\nu$ greater than a certain
critical value $N^{\rm crit}_\nu$, the value of which depends on
the allowed parameter range in the new theory.
Thus, for this new theory, all of the distribution in $N_\nu$
below $N^{\rm crit}_\nu$ must be considered unphysical.
New, obviously relaxed,
bounds on the model parameters follow from application of the Bayesian
method \cite{PDG} of cutting the unphysical region of the parameter
space and renormalizing the remaining distribution to give approximate
(1-$\alpha$) \% CL limits on parameters.\footnote{
We note, but ignore in the following the slight complication
that the bound imposed on model parameters by an (1-$\alpha$) \% CL Bayesian
limit on $N_\nu$ does not exactly correspond to a Bayesian (1-$\alpha$) \%
CL limit imposed directly on the parameter space. This is analogous to
the case of deducing neutrino mass bounds from decay experiments \cite{PDG},
where it is observed that the bound on $m$ is not the same as the
root of the bound derived for $m^2$.}

In this approach, for the case of stable massive neutrinos, we must use
the existing laboratory bounds to the extent that {\em i)} there are
exactly three light neutrinos (LEP) \cite{PDG} and {\em ii)} their
masses are further restricted to $\mnm < 160$ KeV \cite{Lab1} and
$\mnt < 24$ MeV \cite{Lab0}. Then {\em iii)} detailed  nucleosynthesis
computations show us that within these restricted ranges the prediction
for helium abundance always corresponds to having $N^{\rm eff}_\nu
\geq 3$, whence the unphysical region is determined to be
$N_\nu < N^{\rm crit}_\nu =3$.
For example, it has been noted in ref.\ \cite{OS} that `strict' bound
of $N_\nu < 3.13$ based on (\ref{Nlim1}), relaxes to a Bayesian bound
$N_\nu < 3.6$ with $N^{\rm crit}_\nu =3$.  The more of the distribution
lies inside the physical region, the closer the `strict' and Bayesian
bounds come to each other. For example the result (\ref{Nlim2})
implies a `strict' bound of $N_\nu < 4.0$ and, to this accuracy,
is equivalent to the Bayesian  $N^{\rm crit}_\nu = 3$ limit.

After this rather detailed account on how the bounds arise from
the nucleosynthesis, we wish to stress again that, up to the  caveat
mentioned in the footnote 2 in case of the Bayesian approach, the
computation of the nucleosynthesis predictions for a given set of model
parameters on one hand, and finding and imposing the observationally
derived constraints upon them on the other, are {\em unrelated} matters.
The former can be computed exactly, while one's ignorance on the latter
can be parameterized with $N_\nu$.

%

\section {Majorana case}

We now explicitly develop and solve the evolution equations (\ref{geneq})
for the case in which neutrinos are Majorana particles.
 We will take the electron
neutrino to be massless and let the masses of the muon and tau neutrinos
vary freely, keeping in mind however, the laboratory limits $m_\num <
160$ keV \cite{Lab1} and $\mnt < 24$ MeV \cite{Lab0}. We will assume that
electrons and photons are in complete thermal equilibrium and write down
an equation network comprising all neutrinos. Given the assumptions
explained in the previous section, we have
\beqa
\dot n_\nut + 3Hn_\nut & = &  \, \sum_{\alpha=e,\nu_i \neq \nut}
       C(\nut\bnut \leftrightarrow \alpha\bar\alpha ) \nn \\
\dot n_\num + 3Hn_\num & = &  \, \sum_{\alpha=e,\nu_i \neq \num}
       C(\num\bnum \leftrightarrow \alpha\bar\alpha ) \nn \\
\dot n_\nue + 3Hn_\nue & = &  \, \sum_{\alpha=e,\nu_i \neq \nue}
       C(\nue\bnue \leftrightarrow \alpha\bar\alpha ) \nn \\
               &\vdots & ,
\label{majnet}
\eeqa
where we used the compact notation (\ref{lhs}) when writing the left hand
side of the equations and the dots refer to the equations
(\ref{tneu}) and (\ref{ttemp}). In practice, we have to isolate the derivatives
$\dr z_i/\dr \tg$ on the left hand side, as (\ref{majnet}) is truly a network
to solve for the evolution of $z_i$'s. It would not be practical to show the
complicated forms here, however a generic collision term appearing on
the right hand sides of (\ref{majnet}) is given by:
\beqa
  C(\nu_\beta {\bar \nu}_\beta \leftrightarrow \alpha\bar\alpha )
   &\equiv&
    \frac{1}{512\pi^6}(e^{2z_{\nu_\beta}} - e^{2z_\alpha}) \times \nn \\
   &\phantom{=}&
    \times \int {\cal D}\Phi_{\{z_i\}} \int_0^{2\pi} \dr \phi
          \sum_{\rm spin} \mid {\cal M}(\nu_\beta\bar\nu_\beta
             \rightarrow \alpha\bar\alpha )\mid^2 \; S_{\rm in}S_{\rm fi},
\label{majcol}
\eeqa
where the symmetry factors $S_{\rm in}$ and $S_{\rm fi}$, which are equal to
unity for the present case, are included for completeness. We defined a
shorthand notation for the phase space factors
\beqa
 \int {\cal D}\Phi_{\{z_i\}} &\equiv&
      \int_0^\infty \dr k_1
         \int_0^\infty \dr k_2
         \int_{-1}^1 \dr \cos \theta
         \int_{E_{\rm min}}^{E_{\rm max}} \dr E_{p4}
              \frac {k_1^2k_2^2}{\kappa E_{k1}E_{k2}} \times  \nn \\
                   &\phantom{=}& \times \;\; e^{\beta (E_{k1}+E_{k2})}
               f(k_1,z_1)f(k_2,z_2)f(p_3,z_3)f(p_4,z_4),
\label{phasespace}
\eeqa
where $\theta$ is the angle between the incoming 3-momenta ${\bf k}_1$ and
${\bf k}_2$, $\phi $ is the acoplanarity angle between the planes of
incoming and outgoing momenta and
$\kappa \equiv (k_1^2 + k_2^2 - 2k_1k_2\cos\theta )^{1/2}$.
The integration limits in the energy of the outgoing particle are
\beq
E_{\rm \stackrel{\scriptstyle min}{max}} =
         (E_{k1}+E_{k2})\frac{s+m^2_3-m^2_4}{2s}
           \mp \kappa \; \frac{\lambda^{1/2}(s,m^2_3,m^2_4)}{2s},
\label{Elimits}
\eeq
where $\lambda (x,y,z) \equiv (x-y-z)^2-4xy$ and s is the usual invariant
$s=(k_1+k_2)^2$. The generic matrix element is (we always define the
momentum labeling as $(12\rightarrow 34)$ in our matrix elements)
\beqa
  \sum_{\rm spin} \left| {\cal M}(\nu_\beta\bar\nu_\beta \rightarrow
                                \alpha\bar\alpha) \right|^2
 &=&      64 G_F^2 \left\{ (c_{V\alpha }^2 + c_{A\alpha }^2)
                   \left( (k_1\cdot p_4)^2 + (k_2\cdot p_4)^2
                  - m_{\nu_\beta}^2 p_3\cdot p_4  \right)
                   \right. \nn \\
 & & \phantom{Ha }
                   \left. + \;\: (c_{V\alpha }^2 - c_{A\alpha }^2) \,
                        m_\alpha^2 \: \left( k_1\cdot k_2
                                             - 2m_{\nu_\beta}^2 \right)
                   \right\},
\label{matrix}
\eeqa
where the normalization of the couplings is such that for neutrinos $c_{V\nu }
=c_{A\nu }=1/2$ and for electrons $c_{Ve} = 2\sin^2\theta_{\rm W}-1/2$ and
$c_{Ae} = -1/2$, except in the scattering $\nue\bnue \rightarrow e\bar e$,
where due to the additional $W$-channel, effectively
$c_{Ve} \rightarrow 2\sin^2\theta_{\rm W}+1/2$ and $c_{Ae} \rightarrow 1/2$
after a Fierz transformation.
While the matrix element (\ref{matrix}) itself is invariant,
the phase space distributions are written in the rest frame of the
plasma and hence we cannot use the
simple CM-frame expression for $|{\cal M}|^2$.
Of course, in the approximation where one neglects the final state
Pauli blocking factors, phase space integral reduces to an 1 dimensional
integral over an invariant cross section \cite{GG}.  Here we will instead
write the dot-products in the frame specified in the appendix A and perform
the phase space integral without further approximations.

The numerical solution of (\ref{majnet}) proceeds as follows.
For a given (pair of) neutrino mass(es) we begin with
equilibrium distributions at some sufficiently high temperature, in
practice at $T_{\rm init} = 100$ MeV,
and integrate the equation network down until $T_{\rm end} = 1$ KeV
(in the photon temperature), when nucleosynthesis is
essentially over, tabulating the functions $z_i(\tg)$ and the time
temperature relation $t(\tg)$ along the integration. Then the resulting
interpolation tables are used as an input for
a properly extended standard nucleosynthesis code, which we again run
for each mass pair generating isocontours in the primordial helium
abundance.  As described in the previous section, we map the deviation in the
helium abundance to a deviation in the number of neutrino degrees of
freedom: $\dNu= N_\nu -3$. Note that the most natural bound is in fact in
terms of the helium abundance itself, but we are yielding to what has
become a the common practice in expressing the nucleosynthesis bounds.
\begin{figure}
\hspace{2.5truecm}
\epsfysize=8.5truecm\epsfbox{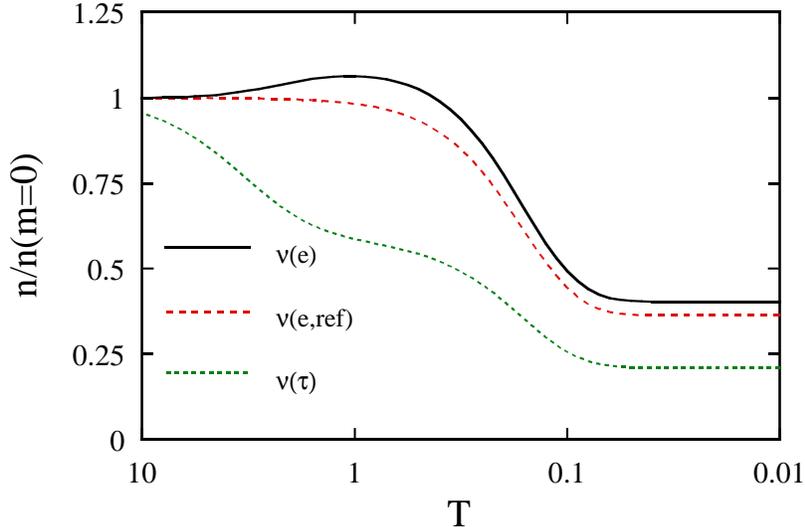}
\vspace{-1truecm}
\baselineskip 17pt
\caption { Shown is the electron neutrino number density
corresponding to the $\nut$ mass of 5 MeV and normalized to equilibrium
density $n_0 \equiv (3\zeta(3)/2\pi^2)\tg^3$. The dashed line shows for
comparison the unperturbed electron neutrino temperature, and the
short dashed line corresponds to the $\nut$ number density. The overall
decrease in the densities around $\tg \sim  0.2$ MeV follows from
electron-positron annihilations which increase $\tg$ relative to
$T_{\nu_e}$.}
\end{figure}
\baselineskip 20pt

In figure 1, we plot a specific run showing the change in the
electron neutrino density due to the annihilation of heavy tau neutrinos
with a mass $\mnt = 5$ MeV.
Because electron neutrinos freeze out at the temperature  $T_{\rm dec}
(\nue ) \simeq 2.3$ MeV, their number density remains close
to the equilibrium value until a few MeV. However, since
the annihilation of $\nut$'s is still occurring below that temperature,
there is a slight increase in the electron neutrino abundance. The excess
is about 10\% at $\tg = 0.7$ MeV, which roughly corresponds to
the temperature when $n/p$-ratio freezes out. The effect of this excess
is to keep $n/p$-ratio in equilibrium until a little later thereby
{\em decreasing} the amount of neutrons and hence the eventual helium
abundance; numerically, in the conventional units of $\dNu$ this effect
is found to correspond to  $\simeq -4.6 \delta n_{\nu_e}$ \cite{EKT}, where
$\delta n_{\nu_e}\sim 0.1$ is the actual change in the electron neutrino
number density (normalized to $n_{\rm eq}= 1$) in the present example.
Combined with the opposite effect on the helium abundance
due to simultaneous slight increase in the energy density, the full effect
of the variation in the electron neutrino density in the present example
is to produce an effective negative contribution of $\dN_\nu^{(\nue )}
\sim -0.36$ to the number of effective species; this example shows the
potential importance of accounting for the electron neutrino abundance
when computing the eventual bounds on masses.

We have computed the total number of effective neutrino species as a
function of neutrino mass, either that of $\num$ or $\nut$, keeping the other
neutrinos massless and display the results in figure 2. We also show
the result for the case where we neglect the effect on the electron
neutrino density. The effect of electron neutrinos is large in the
$few$ MeV region, and it does not affect the eventual bounds
for the small values of $\dNu$ very much. However, for larger values of
$\dNu$ the effect can be significant.

Including the neutrino heating, we find the following bounds on the
masses as a function of $\dNu$-bound: in the small mass side
\beqa
 m^M_\nu/{\rm MeV}
          &<&  ( 0.35\dN^{1/2}_\nu + 0.05\dNu + 0.59\dN^{3/2}_\nu )
              \; \theta (0.15-\dNu) \nn \\
          &+&  (0.09 + 0.47\dNu      + 0.83\dN^2_\nu
                     - 0.72\dN^3_\nu + 0.26\dN^4_\nu )
              \; \theta (\dNu-0.15) \nn \\
         &\equiv & \phi_M(\dNu ),
\label{majsbnd}
\eeqa
where $m_\nu$ is measured in MeVs and $\theta (x)$ is the step function.
This bound is valid for both the muon and the tau neutrino and the
error of the fit is less than 1 per cent for $0.01 < \dNu < 2$.
In the large mass side:
\beqa
m^M_\nu/{\rm MeV} &>&  67.9 - 63.5\dNu      + 38.7\dN^2_\nu
                            - 15.2\dN^3_\nu + 2.4 \dN^4_\nu \nn \\
        &\equiv & \Phi_M(\dNu ),
\label{majlbnd}
\eeqa
which is accurate to better than 1 per cent up to $\Delta N_\nu = 2.5$.
For the both small and large mass limits the dependence on $\eta$ is
of the order of one per cent for $\eta_{10} = 1.4 - 3.8$.

\begin{figure}
\hspace{2.5truecm}
\epsfysize=8.5truecm\epsfbox{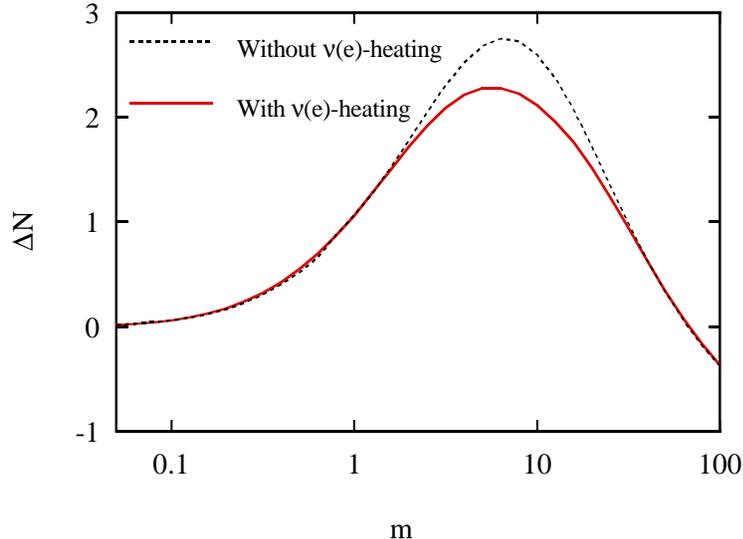}
\vspace{-1truecm}
\baselineskip 17pt
\caption {Shown is the effect of a massive neutrino to
the nucleosynthesis as a function of its mass, expressed in terms of the
effective neutrino degrees of freedom. Dotted line for the comparison
shows the same result ignoring the effect on the electron neutrino
density.}
\end{figure}
\baselineskip 20pt

For example, using the limit $\dNu < 1.0$ from our equation (\ref{Nlim2})
implies the excluded region of 0.93 MeV $< m^M_\nu < 31$ MeV. A more
stringent bound of $\dNu < 0.3$ would have led to bounds 0.31 MeV
$< m^M_\nu < 52$ MeV.
On the other hand, given the present upper laboratory limit of $\mnt < 24$
MeV \cite{Lab0}, opening up a window for a particle in the MeV range would
require a nucleosynthesis bound weaker than $\dNu > 1.3$.
Even with the considerably relaxed nucleosynthesis constraints
obtained neglecting the D and \he3 data \cite{FO2}, this does not
seem very likely. Hence the nucleosynthesis bound is still to be
viewed very much complementary to the laboratory bounds, excluding a
stable Majorana neutrino with a mass in excess of few hundred KeV.
Of course the upper limit found in equation (\ref{majsbnd}) has no
relevance for $\num$ for which the laboratory bound is $\mnm < 160$ KeV
\cite{Lab1}. Moreover, the lower bound coming form the nucleosynthesis
can only be competitive with the laboratory bound if $\dNu < 0.13$.

We complete the Majorana section by noting that the nucleosynthesis
bounds are cumulative; considering the effect of both masses together
yields stronger constraints.  We have computed these bounds by allowing
both neutrinos be massive simultaneously in our code. We found that the
effect in the $\num \leftrightarrow \nut$ reaction rates due to their
having simultaneously nonzero masses is small. Hence the common bounds
can be directly  derived from (\ref{majsbnd}-\ref{majlbnd}). For example
the low mass limit for $\mnt$ in (\ref{majsbnd}) becomes
\beq
\mnt < \phi_M (\dNu - \phi^{-1}_M(\mnm )).
\label{commonbnd}
\eeq
A fit for the inverse function $\phi^{-1}_M(x)$, is explicitly given in
equation (\ref{phi}) in the section 5.2 below.  Similar expression
applies for the large $\mnt$ bound with $\phi_M$ (but not its inverse)
replaced by $\Phi_M$ in (\ref{commonbnd}). The relative error of these
approximate bounds is found to be $\lsim 5$\%.
%

\section {Dirac Case}

In the case of a Dirac neutrino, one is faced with an extra complication
resulting from the chirality of weak interactions; except in the very
nonrelativistic limit, different helicity states have different
interaction strengths. For $m \sim few$ MeV, neutrinos indeed
decouple while semi-relativistic, and it behooves us to write independent
evolution equations for the two helicity populations.
To compare the full treatment to the usual approach which does not
differentiate between the helicities and using the averaged
interaction amplitudes, we note the following:  First, the
R-helicity population interacts much more weakly, and hence their
relic density gets underestimated in the na\"{\i}ve approach. Secondly,
the L-helicity population interacts more strongly than is
assumed in the helicity averaged approach, and there is a compensating
overestimation of their density. Thirdly, the situation is made more
complicated by existence of $t$-channel helicity flipping interactions
that mix the two species.  Clearly, in order to obtain high accuracy
results, a quantitative computation is required to sort out which
of these effects is dominant.

Another problem is that the thermal averaging is more subtle in the
Dirac case, because the spin dependent matrix elements are not Lorentz
invariant. To see this explicitly, consider neutrino-neutrino scattering:
in na\"{\i}ve approach, where one boosts the matrix
element to the CM-frame, it is (see eg.\ (\ref{matrix1})) of the
order $\sim m^4_\nu/E^2$.  This suppression is particular to the CM-frame
however, and the true thermal average is in fact of the same
order $\sim m^2_\nu$
as the other interactions that dominate in the
na\"{\i}ve approach \cite{DKR}.

The technical difficulty is greatly increased
by the very large number of interaction diagrams, in particular because
of a large number of spin flipping $t$-channel processes that
were naturally absent in the Majorana case.
Finally, the lower bounds on the masses
in the Dirac case are sensitive to QCD phase
transition temperature \cite{DKR} because the bounds, as we shall see
is true even for rather large $\dNu$, are saturated by an out of
equilibrium excitation of right handed species below $\Tqcd$,
the excitation process being the more effective the higher $\Tqcd$ is.
We therefore consider large and small mass cases separately

\subsection {Large mass region}

We first concentrate on the large mass region $m \gsim {\cal O}(1)$ MeV.
The distinguishing feature here is that the particles are heavy enough
to have become into equilibrium below the QCD phase transition temperature
so that their distributions can be described by our kinetic equilibrium
approach. The region $m \lsim {\cal O}(1)$ MeV is not well described by
the equations below and we shall return to this point later
(\S \ref{sec:lomass}). The complete
equation network can now be written in the following form
%
\def\Cannm{{
 \sum_{\stackrel{\scriptstyle \lambda = -,+} {\alpha = e,\nue,\num}}
   C(\nutm \bnutl \leftrightarrow \alpha\bar\alpha )}}
\def\Cannp{{
 \sum_{\stackrel{\scriptstyle \lambda = -,+} {\alpha = e,\nue,\num}}
   C(\nutp \bnutl \leftrightarrow \alpha\bar\alpha )}}
%
%
\beqa
\dot n_{\nut_-} + 3Hn_{\nut_-} & = &  \,
        \Cannm + C_{\rm flip}^\tau \nn \\
\dot n_{\nut_+} + 3Hn_{\nut_+} & = &  \,
        \Cannp - C_{\rm flip}^\tau \nn \\
\dot n_\num + 3Hn_\num & = &  \, \sum_{\alpha=e,\nu_{i\lambda} \neq \num}
        C(\num\bnum \leftrightarrow \alpha\bar\alpha ) \nn \\
\dot n_\nue + 3Hn_\nue & = &  \, \sum_{\alpha=e,\nu_{i\lambda} \neq \nue}
        C(\nue\bnue \leftrightarrow \alpha\bar\alpha ) \nn \\
                     &\vdots& \; ,
\label{Dnet}
\eeqa
where the dots again refer to equations (\ref{tneu}) and (\ref{ttemp}).
We included different helicity species only for tau neutrinos, because
in light of the restrictive laboratory bound on muon neutrino, it does not
make sense of computing BBN bounds for $\num$ in the large mass region.
The spin flip terms appearing (\ref{Dnet}) are given by
\beqa
C_{\rm flip}^\tau
&=& \sum_{\alpha = e,\nue,\num}
    \{ \; C(\nutm \alpha \leftrightarrow \nutp \alpha )
        + C(\nutm \bar\alpha \rightarrow \nutp \bar\alpha ) \; \}
\nn \\
&+& \phantom{l}
    \sum_{\lambda = -,+}
    \{ \; C(\nutm \nutl \leftrightarrow \nutp \nutl )
        + C(\nutm \bnutl \leftrightarrow \nutp \bnutl ) \; \}
\nn \\
 &+& 2 C(\nutm \nutm \leftrightarrow \nutp \nutp ),
\label{flips}
\eeqa
where the factor of 2 in the last term accounts for the fact that this
interaction changes the $\nu-$number by 2 units.
Generic collision terms appearing in definitions (\ref{Dnet} -
\ref{flips}) are defined  and normalized similarly to the equations
(\ref{majcol}-\ref{Elimits}). For example (from now on we will denote
$\nutl$ by $\nul{}$).
\beqa
  C(\nul{1} \bnul{2} \leftrightarrow \alpha\bar\alpha )
  &\equiv&
   \frac{1}{512\pi^6}(e^{z_{\lambda 1} + z_{\lambda 2}}
                    - e^{2z_\alpha}) \times \nn \\
  &\phantom{=}&
   \times \int {\cal D}\Phi_{\{z_i\}} \int_0^{2\pi} \dr \phi
              \sum_{\rm spin} \mid {\cal M}(\nul{1}\bnul{2}
               \rightarrow \alpha\bar\alpha )\mid^2,
\label{dircol}
\eeqa
where we dropped the symmetry factors which equal to unity and the
annihilation matrix element is given by
\beqa
    \sum_{\rm spin} \mid {\cal M}(\nul{1}\bnul{2}
               \rightarrow \alpha\bar\alpha )\mid^2
 &=&
       16G_F^2  \left\{ (c_{V\alpha }^2 + c_{A\alpha }^2)
                   \left( K_{\lambda 1}\cdot p_3
                          K_{-\lambda 2}\cdot p_4 \right)
                \right. \nn \\
 & & \phantom{Han }
                \left.  - \frac{1}{2}
                          (c_{V\alpha }^2 - c_{A\alpha }^2) \,
                          m_\alpha^2 \: K_{\lambda 1}\cdot K_{-\lambda 2}
                \right\}.
\label{matrix1}
\eeqa
Here, in order to condense the notation, we dropped the terms
that vanish, and combined others that are equal under the integration;
similar simplifications are made in other matrix elements
following below.
The coefficients $c_{V\alpha }$ and $c_{A\alpha }$ have been defined in
section 4 and the 4-vector $K^\mu_{\lambda} \equiv k^\mu -
m_\nu s^\mu_{\lambda}$ is related to the `spin vector'
$s^\mu_{\lambda}$ of $i$th neutrino and can be written as
\beq
K^\mu_{\lambda} = (E-\lambda (E^2 - m^2_\nu)^{1/2})
                    (1;-\lambda {\bf k}/k).
\label{kmu}
\eeq
Other matrix elements appearing in the collision terms above include
the $t$-channel scattering off the electrons and other neutrinos and
their antiparticles.  Under the assumption that the chemical potentials
are small ($\ll 1$) the distribution functions for particles and
antiparticles are equivalent, and we can add their contributions under
the integral:
\beqa
    \sum_{\rm spin,\; \beta = \alpha, \bar\alpha}
              \mid {\cal M}(\nul{1}\beta
               \rightarrow \nul{2} \beta )\mid^2
 &=&
       16G_F^2  \left\{ (c_{V\alpha }^2 + c_{A\alpha }^2)
                   \left( K_{\lambda 1}\cdot p_2 K_{\lambda 1} \cdot p_4
                        + K_{\lambda 1}\cdot p_4 K_{\lambda 3} \cdot p_2
                   \right)  \phantom{\frac{1}{2}}
                \right. \nn \\
 & & \phantom{Ham }
                \left.  - (c_{V\alpha }^2 - c_{A\alpha }^2) \,
                           m_\alpha^2 \: K_{\lambda 1}\cdot K_{\lambda 3}
                \right\}.
\label{matrix2}
\eeqa
Finally, self scatterings can all be derived from the matrix elements
\beq
  \mid {\cal M}(\nu_{\lambda 1}\bar \nu_{\lambda 2}
    \rightarrow \nu_{\lambda 3}\bar \nu_{\lambda 4} )\mid^2
\; = \; 8G_F^2 \; K_{\lambda 1}\cdot K_{-\lambda 4}
                  K_{\lambda 2}\cdot K_{-\lambda 3}
\label{matrix3}
\eeq
and
\beq
  \mid {\cal M}(\nu_{\lambda 1}\nu_{\lambda 2}
    \rightarrow \nu_{\lambda 3}\nu_{\lambda 4} )\mid^2
\; = \; 8G_F^2 \; K_{\lambda 1}\cdot K_{\lambda 2}
                  K_{\lambda 3}\cdot K_{\lambda 4}.
\label{matrix4}
\eeq
The symmetry factors appearing in the collision integrals are
equal to one everywhere except the reactions
corresponding to (\ref{matrix4}). There, the symmetry factor is one half,
because of the degeneracy in either initial or in final state, except
for the reaction $\nutm\nutm\rightarrow \nutp\nutp$, where the symmetry
factor is $1/4$. Note that this
reaction changes $\nutm$ number by two units,
but that was explicitly taken into account in the equation (\ref{flips}).

The collision integrals of the massless $\nue$ and $\num$ appearing
in (\ref{Dnet}) are obtained from the Majorana matrix element
(\ref{matrix}) in the limit $m_\nu = 0$.
Equations (\ref{matrix},\ref{matrix1}-\ref{matrix4})
exhaust the list of interactions relevant for
the evolution of the neutrino ensembles. Each of the matrix elements
(\ref{matrix}, \ref{matrix1}-\ref{matrix4}) is a polynomial at
most of second order of the cosine of the acoplanarity $\phi$.
We integrate over $\phi$ analytically, after which the remaining
4-dimensional integral is performed numerically using the special
frame introduced in the appendix A.
In figure 3 we show the temperature dependence of the annihilation
and flip rates $\Gamma_\nul{\pm}$ and $\Gamma_{\rm flip}$, defined by
(see the appendix A) $\Gamma_i = \sum_j n_j \vs {(ij\rightarrow kl)}$
for the particular case of $\mnt = 5$ MeV along with the Hubble
expansion rate. One sees how the
right helicity population drops from the equilibrium much earlier than
left helicity states.  Yet both states are in complete equilibrium
until well below the QCD phase transition temperature
$\Tqcd \sim few$ 100 MeV.

\begin{figure}
\hspace{2.5truecm}
\epsfysize=8.5truecm\epsfbox{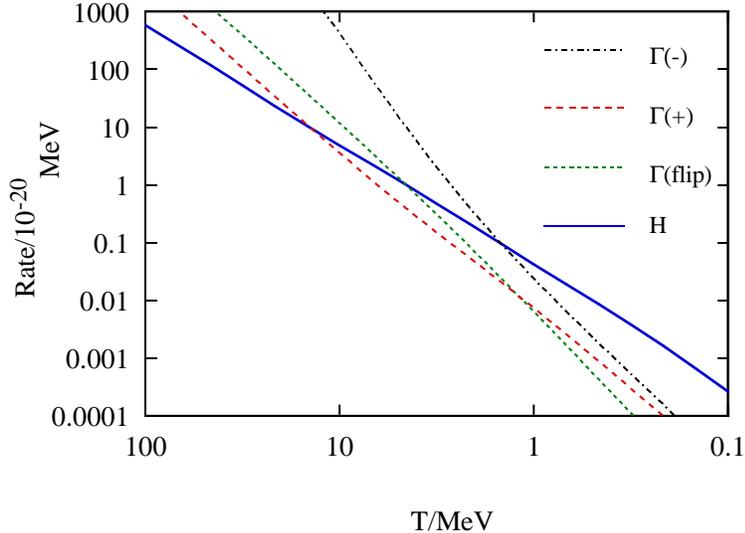}
\vspace{-1truecm}
\baselineskip 17pt
\caption {Shown are the total annihilation rates $\Gamma_{\nul {\pm}}$
and the total flip rate $\Gamma_{\rm flip}$ (see the text) in comparison
with the Hubble expansion rate $H(T)$.}
\end{figure}
\baselineskip 20pt
In figure 4, we show a particular example of the evolution of the
neutrino energy densities as a function of time.
While negative and positive helicity populations have
different densities from each other, their average comes
 close to that obtained in the helicity averaged approach.

\begin{figure}
\hspace{2.5truecm}
\epsfysize=8.5truecm\epsfbox{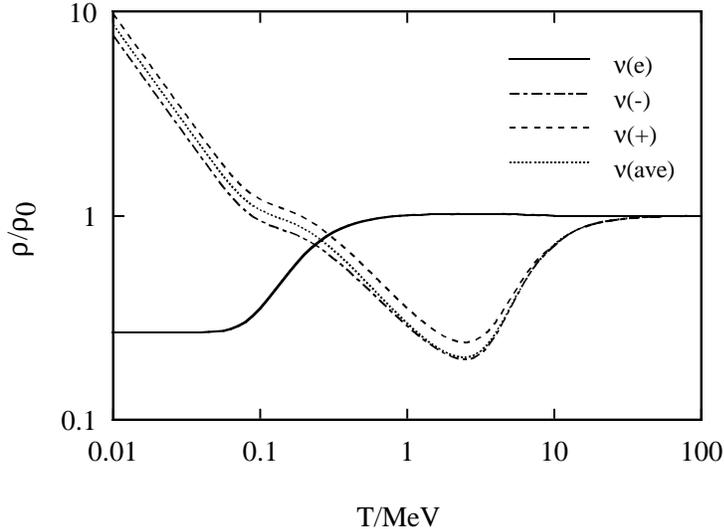}
\vspace{-1truecm}
\baselineskip 17pt
\caption {Shown is the evolution of the energy densities of $\nutm$,
$\nutp$ and $\nue$ corresponding to a run of the code with $\mnt =
20$ MeV and normalized to $\rho_0 = 7\pi^2T^4/240$. Also shown is
the energy density corresponding to $\nut$ in the helicity averaged
approximation.}
\end{figure}
\baselineskip 20pt

In figure 5, we show the change in the helium abundance in the
Dirac case for $m^D_\nut \gsim 1$ MeV.  We did our computations also
using the helicity averaged approach. The final results turned out
to be very close to the full solution, in particular in the region of
interest for nucleosynthesis bounds.  While one might have expected this
result on qualitatively, the quantitative proof only could follow from
a numerical calculation.

We find that the nucleosynthesis bound on the $\nut$ mass is fitted to an
accuracy of one per cent in the range $0<\dNu<2.5$ by:
\newpage
\beqa
m^D_\nut/{\rm MeV} &<& 37.8 - 26.9\dNu      + 21.3\dN^2_\nu
                          - 15.5\dN^3_\nu +  6.3\dN^4_\nu
                          -  0.1\dN^5_\nu \nn \\
                 &\equiv & \Phi_D(\dNu ).
\label{diracbnd}
\eeqa
This is the main result of this subsection.  One observes that the
nucleosynthesis constraint allows a stable Dirac neutrino just below
the laboratory bound $\mnt < 24$ MeV, given that $\dNu > 0.8$. This
seems to be admissible given the relaxed constraint following from
the equation (\ref{Nlim2}); indeed, the bound $\dNu < 1.0$ gives
the constraint $m^D_\nut > 22$ MeV.  More stringent bound of $\dNu
< 0.3$ would lead to  $m^D_\nut > 31$ MeV. Our result (\ref{diracbnd})
differs noticeably from the previous computations and fall
roughly in between the results obtained in \cite{kolb} and \cite{DR}
given the particular values for the bound for $\dNu$ used therein.

\begin{figure}
\hspace{2.5truecm}
\epsfysize=8.5truecm\epsfbox{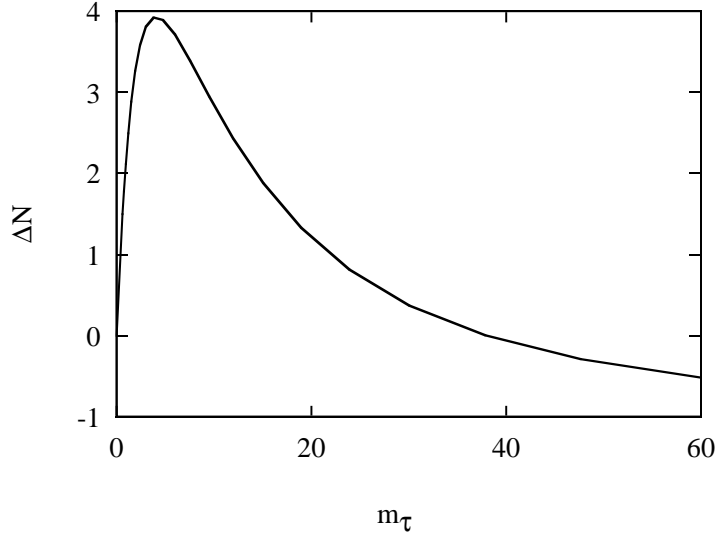}
\vspace{-1truecm}
\baselineskip 17pt
\caption {Shown is the effective number of degrees of freedom for
a Dirac tau neutrino.}
\end{figure}
\baselineskip 20pt
%

\subsection {Small mass region}
\label{sec:lomass}

The mass of a neutrino is considered `small' if the R-helicity population
is out of equilibrium below $\Tqcd$.  Even in this case however,
a significant amount of R-helicity states may be produced by out of
equilibrium spin flip processes \cite{DKR}.
We shall see below that even for very large $\dNu$, the lower limit
in the exclusion region indeed is saturated by a mass for which R-helicity
states are out of equilibrium.

In the small mass region it does not make sense to describe the R-helicity
system with the distribution function (\ref{fansaz}). Instead, noting
that the R-helicity states are produced from an equilibrium ensemble
of left helicity states through  very mildly energy dependent spin
flip interactions, the increments in R-helicity population appear with a
local (in time) equilibrium density characterized by the excitation
temperature $T(t_{\rm ex })$.  Provided that the backscattering is not
very efficient (to be checked {\em a posteriori}), one can compute
the total energy density in the right handed species very accurately
by including the dilution due to entropy production subsequent to
excitations. This program was carried out in the reference \cite{DKR}.
Here we stress that the underlying assumption of no backscattering and
hence the bounds are valid for surprisingly large values of $\dNu$.
To this end we extend the treatment of \cite{DKR} by including a simple
but accurate model of backscattering.  We also point out and correct
an inaccurate treatment of the effect of a mass of a neutrino for the
nucleosynthesis in the final stages of the analysis of \cite{DKR}.
The corrected analysis turns out to give bounds roughly 30 per cent more
stringent than those of ref.\ \cite{DKR}.

We will write down the starting point of our analysis using the results
obtained in ref.\ \cite{DKR}. Because the spin flip processes involving
only one right helicity state at the time are by far the dominant
interactions here, an equation which simply models the back
scatterings can be written as
\beq
\dot \rho_\nu + 4H\rho_\nu
=  \frac{G_F^2m_\nu^2}{2\pi^5} {\hat K}_{\rm eff} T^7\;
   {\cal C}_\nu (T) \; ( 1 - \rho/\rho_{\rm eq} )
\label{rhoequ}
\eeq
where $\rho_{\rm eq} = 7\pi^2T^4/240$ is the equilibrium energy density
of massless neutrinos, ${\hat K} \simeq 16.52$ includes counting over
all channels the Fermi-Dirac correction due to the statistics of the
L-helicity particles and the functions ${\cal C}_i$ are given
by \cite{DKR}
\beqa
{\cal C}_\mu (T)&=&
1 + 0.81d_\mu (T)
+ 3.71 \times 10^{-2} \left( (\frac{f_{\pi^0}}{T})^2 z_0^4K_2(z_0) \right.
\nonumber \\ & & \phantom{hanna} \left.
+ (\frac{f_{\pi^\pm}}{T})^2 z_\pm^4(1-y^2)
                \{ (1+y^2)K_2(z_\pm) - y^2K_0(z_\pm) \} \right)
\nonumber \\
{\cal C}_\tau (T) &=& 1 + 0.06d_\tau(T) +  3.71 \times 10^{-2}
 (\frac{f_{\pi^0}}{T})^2 z_0^4K_2(z_0)
\label{calCt}
\eeqa
where the functions $d_i(T)$ express the nontrivial temperature
dependence (deviation from the $T^7$-law) of the scattering collision
term due to the scatterings off muons.  The terms involving
the modified Bessel functions $K_i(z)$ correspond to pion decays with,
$z_0\equiv m_{\pi^0}/T$, $z_\pm\equiv m_{\pi^\pm}/T$, $y \equiv
m_{\pi^\pm}/m_\mu$ and $f_{\pi^0} \simeq 93$ MeV and
$f_{\pi^\pm} \simeq 128.7$ MeV.  Equation (\ref{rhoequ}) is easily
integrated along with the equation (\ref{ttemp}) to yield a double
integral expression for the relative energy density $r\equiv
\rho/\rho_{\rm eq}$
\beqa
r(x) &=&
\int_x^1 \dr x' (\frac{h_I(x)}{h_I(x')})^{4/3} A(x')
\exp \int_x^{x'} \dr x'' A (x'') \nonumber \\
     &+&
\rqcd \; (\frac{h_I(x)}{h_I(x_{\rm QCD})})^{4/3}
\exp \int_1^x \dr x' A (x'),
\label{requ}
\eeqa
where $\rqcd = (17.25/60)^{4/3} \simeq 0.19$ is the diluted energy
density of the equilibrium ensemble of R-helicity population decoupled
above the QCD phase transition, $x \equiv T/\Tqcd$, and
\beq
A (x) \simeq 2.88 \; m_\nu^2\; \T100\;
(\frac{10.75}{g_*(x)})^{1/2} (1+\frac{x}{3h_I}\frac{\dr h_I}{\dr x}) \;
{\cal C}_\nu (x),
\label{psi}
\eeq
where $\T100 \equiv \Tqcd/100$ MeV, $x \equiv T/\T100$ and $m_\nu$ is
in units MeV. Expression (\ref{requ}) obviously reduces to those of
\cite{DKR} when backscattering is neglected.  We computed the relevant
value of the function $r$ during the nucleosynthesis, $r(0)$, for a large
set of parameters $m_\nu$ and $\Tqcd$ and found that it is to a reasonable
approximation
\beq
r(0) \simeq r'_{\rm QCD}e^{-\Delta r_i} + ( 1-e^{-\Delta r_i}),
\label{fitrequ}
\eeq
where $r'_{\rm QCD} \simeq 0.1$ and
\beqa
\Delta r_\mu  = (2.89 + 2.25 \T100 )\; m^2_{\nu_\mu} \nonumber \\
\Delta r_\tau = (1.29 + 1.34 \T100 )\; m^2_{\nu_\tau}
\label{ab}
\eeqa
The accuracy is no worse than 3 per cent for $\Tqcd = 100-200$ MeV
and $\dNu < 1.5$, which corresponds to the whole region of applicability
of the final result. For $\Tqcd = 300-400$ it undershoots by about 10
per cent at large $\dNu \gsim 1.5$.
It should be noted that
due to the mass effects, a single neutrino with an equilibrium density
effectively corresponds to $1+f(m)$ species. Using the nucleosynthesis
code we have computed a fit for this function. For moderately small masses
$m_\nu \lsim 0.6$ MeV, it in fact coincides with the function
$\phi^{-1}(m_\nu )$ for the number of effective degrees of freedom for
the small mass majorana neutrinos, defined in equation
(\ref{majsbnd}):
\beqa
f(m) = \phi^{-1}_M (m)
    &=& (-18.6 m^3 + 7.9m^2 + 0.02m) \; \theta (0.15-m) \nn \\
    &+& (0.007m^4 - 0.019m^3 + 0.237m^2 + 1.40m - 0.09)
         \; \theta (m-0.15).
\label{phi}
\eeqa
The function $f(m)$ defined above differs considerably from the
fit $f(m^2)$ used in its place in ref.\ \cite{DKR}.\footnote{
The fit function $f(m^2)$ used in \cite{DKR} unfortunately
strongly underestimated the effect of a small neutrino mass to the
nucleosynthesis. This is because it apparently failed to correctly
model the dominant source of the effect of a neutrino mass for the
nucleosynthesis; the change in the capture time of free neutrons.
Indeed, at the capture temperature $T_\gamma \sim 0.1$ MeV, the mass
is typically dominating over the radiation, whence one expects a strong
linear correlation between mass and the induced effective chance in
$N_\nu$, as is seen in our fit for $f(m)$ above.}
Using the approximation (\ref{fitrequ}) we are finally led to the
constraint equation
\beq
f(m)+ (1 + f(m))(r'_{\rm QCD}e^{-\Delta r_i} + (1-e^{-\Delta r_i}))
< \dNu.
\label{Dlowbnd}
\eeq
We plot the bound for the tau neutrino in figure 6 for
$\Tqcd = 100-400$ MeV with our improved fit function $f(m)$ and using the
exact expression (\ref{requ}) for $r(0)$. We also show the value of $r(0)$
to underline how the effective value for $\dNu$ greatly exceeds
$r(0)$ for even moderate masses. This is the reason why the
backscattering correction is so small (we find it is typically
at most 10 per cent for $\dNu \lsim 1.5$).

\begin{figure}
\hspace{2.5truecm}
\epsfysize=8.5truecm\epsfbox{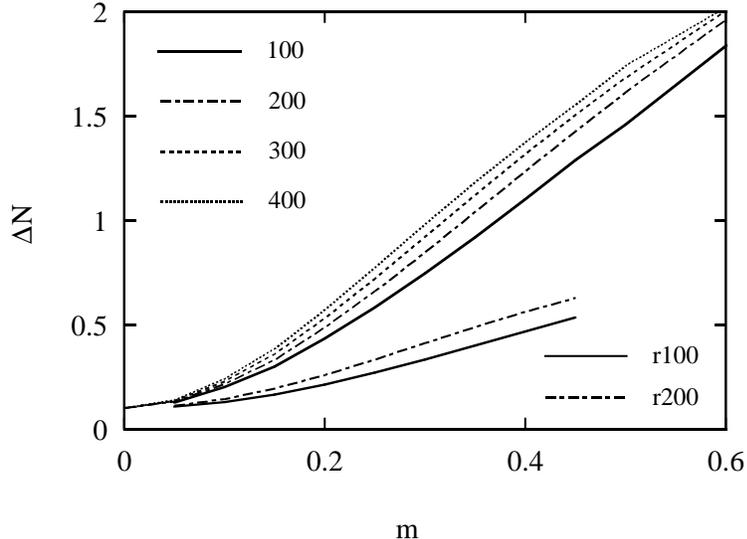}
\vspace{-1truecm}
\baselineskip 17pt
\caption {Shown is the bound on the Dirac $\nut$ mass as a function
of the nucleosynthesis constraint $\dNu$ for $\Tqcd = 100-400$ MeV
The two short lines show the function $r(0)$ for $\Tqcd = 100 - 200$
MeV.}
\end{figure}
\baselineskip 20pt

There is a slight inaccuracy in the derivation above that goes over
the stated approximations. Namely, we have computed the mass effect
of the excited right handed population by multiplying by $f(m)$ their
fraction of the energy density $r$. It is clear however that this
mass effect actually depends on the relative {\em number} density $n$
instead. To correct for this inaccuracy would require computing also
$n$ in a way similar we found $r$ above. This would merely require
re-evaluating the collision terms and correcting the power for the
dilution factor $4/3 \rightarrow 1$ in equation (\ref{requ}). We
will not do so here, because the error made is very small. Indeed, using
the equation (\ref{fitrequ}) it is easy to show that the relative error
(undershoot) in $\dNu$ is generously bracketed by $\delta (\dNu ) <
((a-1)r'_{\rm QCD}e^{-\Delta r} + (b-1)(1-e^{-\Delta r}))f(m) \lsim
0.04\dNu$, in the mass range of interest ($a\equiv (60/10.75)^{1/3}$
and $b\equiv (17.25/10.75)^{1/3}$). Moreover, this error tends to cancel
the error made by the approximation (\ref{fitrequ}).

In case of the muon neutrino, our bound becomes competitive with
the laboratory bound of $\mnm < 160$ KeV, when $\dNu < 0.39$ (0.44)
for $\Tqcd = 100$ (200) MeV.  Since we are finding that the
function $f(m)$ was inaccurately estimated in ref.\ \cite{DKR} we
give for comparison the correct bounds for the values of $\Tqcd$
and $\dNu < 0.3$ discussed there:
\beq
\label{eq:bounds}
m_{\nu_\mu} \: \lsim \:
\left\{
\begin{array}{lll}
&130~{\rm KeV} &
\;  \; T_{QCD}=100~{\rm MeV} \\
&120~{\rm KeV} &
\;  \; T_{QCD}= 200~{\rm MeV} \nonumber
\end{array}
\right. \,
\eeq
\beq
\label{eq:bounds2}
m_{\nu_\tau} \: \lsim \:
\left\{
\begin{array}{lll}
&150~{\rm KeV} &
\;  \; T_{QCD}=100~{\rm MeV} \\
&140~{\rm KeV} &
\;  \; T_{QCD}= 200~{\rm MeV.} \nonumber
\end{array}
\right.
\eeq
For $\dNu < 1.0$, these limits become:
\beq
\label{eq:bounds3}
m_{\nu_\mu} \: \lsim \:
\left\{
\begin{array}{lll}
&310~{\rm KeV} &
\;  \; T_{QCD}=100~{\rm MeV} \\
&290~{\rm KeV} &
\;  \; T_{QCD}= 200~{\rm MeV} \nonumber
\end{array}
\right. \,
\eeq
\beq
\label{eq:bounds4}
m_{\nu_\tau} \: \lsim \:
\left\{
\begin{array}{lll}
&370~{\rm KeV} &
\;  \; T_{QCD}=100~{\rm MeV} \\
&340~{\rm KeV} &
\;  \; T_{QCD}= 200~{\rm MeV.} \nonumber
\end{array}
\right.
\eeq

Before concluding, we discuss the connection between the computations in the
high mass and the low mass regions. It should be obvious that
when the function $r(0)$ is close to 1, one actually enters the
region where the kinetic-equilibrium treatment employed at the
high mass region becomes valid.  However, one would expect that the
connection of the low and high mass solutions is not completely
smooth, because close to the crossing point the right helicity
population is not quite in complete equilibrium, nor completely
out of  equilibrium ($r_{QCD}$ is bigger than assumed in the low
mass treatment).  Hence it is difficult to improve the computation
quantitatively without a detailed knowledge of the QCD phase
transition  dynamics. We conclude that the bound (\ref{Dlowbnd})
should be trusted until about $\dNu \sim 1.0$, above which
there may be large (few tens of per cents) uncontrolled
uncertainties in the results.

%

\section {Conclusions}

In conclusion, we have carefully computed the mass bounds on the stable
Majorana and Dirac neutrinos arising from nucleosynthesis constraints.
We also discussed in detail how nucleosynthesis constraints on particle
physics models arise, and how (and to what extent) they can generically
be modeled through the effective number of degrees of freedom. In our
computation of the mass bounds we included the effects of heating of the
electron neutrino system as a result of the annihilations below the
$\nue$ freeze-out temperature and the effect of chirality in the weak
interactions on the evolution of different helicity components
in the case of a Dirac neutrino. We also computed the bounds for the
case in which both $\num$ and $\nut$ are massive simultaneously, resulting
in stronger constraints.  Most importantly, we computed
all bounds as {\em functions} of the actual nucleosynthesis constraint
on the effective number of neutrino degrees of freedom $\dNu$, except
in the case of light Dirac neutrinos, where nevertheless an implicit
function of a form $\phi_D (\dNu, m_\nu$, $\Tqcd) = 0$ was derived for
the bound. We claim that in all our bounds, the theoretical uncertainty
is negligible and therefore realistic bounds, or estimates of the
uncertainties in the bounds can be obtained solely on the basis of
a separate analysis of the determination of the bound on $\dNu$.

\section* {Acknowledgements}

This research is supported by the DOE grant DE-AC02-83ER40105.
%

\appendix
\section{Phase space integrals}

In this appendix we define the phase space co-ordinate system we employed
in our computations and write down the matrix element in terms of these
variables in a couple examples. We took the
independent variables to be the magnitude of the two 3-momenta of the
incoming states, $k_1$ and $k_2$, energy of one of the outgoing states
$E_{q2}$, the angle between the incoming states $\theta$ and the
acoplanarity angle $\phi$ between the collision planes. In terms of
these variables relevant 4-momenta have the expressions (we are using
the same expression for the 4-momentum and the magnitude of the
corresponding 3-momentum; what is meant in each occasion should be
obvious however)
\beqa
k_1 &=& (E_{k1};  k_1\sin\theta_1, 0, k_1\cos\theta_1) \nn \\
k_2 &=& (E_{k2}; -k_2\sin\theta_2, 0, k_2\cos\theta_2) \nn \\
p_4 &=& (E_{p4};  p_4\sin\theta_K\cos\phi,
                  p_4\sin\theta_K\sin\phi,
                  p_4\cos\theta_K),
\label{fourmom}
\eeqa
where the angles are defined as
\beqa
\cos \theta_1 &=& (k_1 + k_2\cos\theta )/\kappa \nn\\
\cos \theta_2 &=& (k_2 + k_1\cos\theta )/\kappa \nn\\
\cos \theta_K &=& \frac{1}{\kappa Q_1}\left( E_{p4}(E_{k1}+E_{k2})
                   - E_{k1}E_{k2} + k_1k_2\cos\theta
                   \phantom{\frac{1}{2}} \right. \nn \\
              & & \phantom{Hannatyt} \left.
                 - \frac{1}{2}(m_1 + m_2 + m_4 - m_3)
                   \right).
\eeqa
and $\kappa=({\bf k}_1+{\bf k}_2)^2$ (cf.\ equation (\ref{Elimits})).
In terms of these variables the matrix element (\ref{matrix})
in the Majorana case becomes
\beqa
\int\dr\phi |{\cal M}|^2 &=& 128\pi G_F^2
        \left\{
           (c_{V\alpha }^2 + c_{A_\alpha }^2)
             \left( s^2/2
                  - 2s( E_{k1}E_{p4} + k_1p_4\cos\theta_1\cos\theta_K )
                  \right. \right. \nn \\
 & & \phantom{Hannaty }  \left. \left.
                  + 4 E_{k1}^2E_{p4}^2
                  - 8 E_{k1}E_{p4}k_1p_4\cos\theta_1\cos\theta_K
                         \right.\right. \nn \\
 & & \phantom{Hannaty }  \left. \left.
                  + 2k_1^2p_4^2\left( 2\cos\theta_1^2\cos\theta_K^2
                                     +\sin\theta_1^2\sin\theta_K^2 \right)
                  - m_{\nu_\beta}^2( s-2m_\alpha^2 )
             \right)      \right. \nn \\
 & & \phantom{Hanna }    \left.
        +\;(c_{V\alpha }^2 - c_{A_\alpha }^2)
         \, m_\alpha^2 \: \left( s - 6m_{\nu_\beta}^2 \right)
        \right\},
\label{majmat}
\eeqa
where $s = 2E_{k1}^2E_{k2} - 2k_1k_2\cos\theta + m_{i1}^2 + m_{i2}^2$.
We have checked numerically that when integrated over the phase space
in the Maxwell-Boltzmann approximation, the matrix element (\ref{majmat})
reproduces the much simpler thermal average over the invariant cross
section \cite{GG}
\beqa
\vs {} &\equiv & \frac{1}{512\pi^6n_1n_2}
     \times \int {\cal D}\Phi_{\{0\}} \int_0^{2\pi} \dr \phi
          \sum_{\rm spin} \mid {\cal M}(\nu_\beta \bar\nu_\beta
             \rightarrow \alpha \bar \alpha \mid^2 \; S_{\rm in}S_{\rm fi}
    \nn \\
  &\stackrel{\scriptstyle \rm MB}{\mapsto} &
{1\over 8m^4TK_2^2({m\over T})} \int_{4m^2}^\infty
\dr s \sqrt {s}(s-4m^2)K_1({\sqrt s\over T}) \sigma_{\rm CM}(s),
\label{MBlimit}
\eeqa
where the cross section is easily obtained by integrating the matrix
element (\ref{matrix})
\beqa
\sigma_{\rm CM}(s) &=& \frac{G_F^2}{2\pi s}\frac{v_f}{v_i}
                 \left\{ (c_{V\alpha }^2 + c_{A_\alpha }^2)
                      \left( s^2(1+\frac{1}{3}v_i^2v_f^2)
                            - 4m_\nu^2(s-2m_\alpha^2)  \right)
                 \right. \nn \\
                   & &
                 \left. \phantom {\frac{G_F^2}{\pi}a}
                      + 4(c_{V\alpha }^2 - c_{A_\alpha }^2)
                         m_\alpha^2 (s-6m_\nu^2)
                \right\}.
\label{majxsect}
\eeqa
Similar checking is not directly possible in the Dirac case, because there
the matrix element is {\it not} invariant and one does not expect the
helicity amplitudes to reduce to the simple expression (\ref{MBlimit}).
However, the {\it helicity summed} amplitude is of course again an invariant
and we undertook to check in every case separately that when summed over
initial state helicities our thermal averages again did reproduce the
simpler results (\ref{MBlimit}) over the total scattering cross section
in the Maxwell-Boltzmann limit.

%
%
%
%

\end{document}